\documentclass[prl,showpacs,superscriptaddress,twocolumn]{revtex4}

\usepackage{graphicx}

\begin{document}

\title{Entanglement Entropy in Random Quantum Spin-$S$ Chains}
\author{A. \surname{Saguia}}
\email{amen@if.uff.br}
\affiliation{Instituto de F\'{\i}sica - Universidade Federal Fluminense,
Av. Gal. Milton Tavares de Souza s/n, Gragoat\'a, Niter\'oi, 24210-346, RJ, Brazil\vspace{0.1cm}}
\author{M. S. \surname{Sarandy}}
\affiliation{Departamento de Ci\^encias Exatas, P\'olo Universit\'ario de Volta Redonda,
Universidade Federal Fluminense, Av. dos Trabalhadores 420, Volta Redonda,
27255-125, RJ, Brazil \vspace{0.1cm}}
\author{B. \surname{Boechat}}
\affiliation{Instituto de F\'{\i}sica - Universidade Federal Fluminense,
Av. Gal. Milton Tavares de Souza s/n, Gragoat\'a, Niter\'oi, 24210-346, RJ, Brazil\vspace{0.1cm}}
\author{M. A. \surname{Continentino}}
\affiliation{Instituto de F\'{\i}sica -
Universidade Federal Fluminense, Av. Gal. Milton Tavares de Souza
s/n, Gragoat\'a, Niter\'oi, 24210-346, RJ, Brazil\vspace{0.1cm}}

\date{\today }

\begin{abstract}
We discuss the scaling of entanglement entropy in the random
singlet phase (RSP) of disordered quantum magnetic chains of
general spin-$S$. Through an analysis of the general structure of
the RSP, we show that the entanglement entropy scales
logarithmically with the size of a block and we provide a closed
expression for this scaling. This result is applicable for
arbitrary quantum spin chains in the RSP, being dependent only on
the magnitude $S$ of the spin. Remarkably, the logarithmic scaling
holds for the disordered chain even if the pure chain with no
disorder does not exhibit conformal invariance, as is the case for
Heisenberg integer spin chains. Our conclusions are supported by
explicit evaluations of the entanglement entropy for random
spin-$1$ and spin-$3/2$ chains using an asymptotically exact
real-space renormalization group approach.

\end{abstract}


\maketitle

\section{I - Introduction}

\begin{figure}[th]
\centering {\includegraphics[angle=0,scale=0.45]{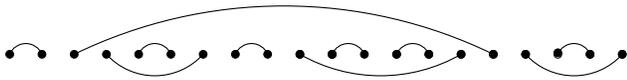}}
\caption{A schematic picture of the RSP. Spin singlets are composed at arbitrary distances.}
\label{f1}
\end{figure}

Quantum information science provides fruitful connections between
different branches of physics. In this context, the relationship
between entanglement, which is a fundamental resource for quantum
information applications~\cite{Nielsen:book}, and the theory of
quantum critical phenomena~\cite{Sachdev:book,Continentino:book}
has been a focus of intensive research. Specifically, entanglement
measures have been proposed as a tool to characterize quantum
phase transitions (See, e.g.,
Refs.~\cite{Osterloh:02,Nielsen:02,Vidal:03,Wu:04,Oliveira:06}).
In this direction, a successful approach has been the analysis of
bipartite entanglement in quantum systems as measured by the von
Neumann entropy. Given a quantum system in a pure state
$|\psi\rangle$ and a bipartition of the system into two blocks $A$
and $B$, entanglement between $A$ and $B$ can be measured by the
von Neumann entropy ${\cal S}$ of the reduced density matrix of
either block, i.e.,
\begin{equation}
{\cal S}=-\textrm{Tr} \left( \rho_A \log_2 \rho_A \right) = -\textrm{Tr} \left( \rho_B \log_2 \rho_B \right),
\label{vonNeumann}
\end{equation}
where $\rho_A=\textrm{Tr}_B \rho$ and $\rho_B = \textrm{Tr}_A
\rho$ denote the reduced density matrices of blocks $A$ and $B$,
respectively, with $\rho=|\psi\rangle\langle \psi|$. By evaluating
${\cal S}$ for quantum spin systems, Ref.~\cite{Vidal:03}
numerically found that entanglement displays a logarithmic scaling
with the size of the block in critical (gapless) chains and
saturates to a constant value in chains with a gap for
excitations. The logarithmic scaling was proven in general for
one-dimensional quantum models exhibiting conformal
invariance~\cite{Korepin:04,Calabrese:04}, with the scaling
governed by a universal factor, given by the central charge of the
associated conformal field theory. Indeed, for a block of spins of
length $L$ in a quantum chain, von Neumann entropy ${\cal S}(L)$
scales as
\begin{equation}
{\cal S}(L) = \frac{c}{3}\, \log_2 L + k \, ,
\label{ScalingOrdered}
\end{equation}
where $c$ is the central charge and $k$ is a non-universal
constant. Recently, the behavior of entanglement entropy in
critical spin chains has also been discussed in presence of
disorder~\cite{Refael:04,Laflorencie:05,chiara,Santachiara:06,Igloi:07}.
Disorder appears as an essential feature in a number of condensed
matter systems, motivating a great deal of theoretical and
experimental research (e.g., see~\cite{Young:98,Igloi:05}).
Moreover, disorder usually introduces a further effect, namely,
the breaking of conformal symmetry in a critical model.
Remarkably, Refael and Moore~\cite{Refael:04} have shown that,
even in the absence of conformal invariance, due to disorder, the
logarithmic scaling given by Eq.~(\ref{ScalingOrdered}) holds for
the spin-1/2 random exchange Heisenberg antiferromagntic chain
(REHAC) and for the Griffiths phase of the random transverse field
Ising chain.  In this case, however, it is governed by an effective
central charge ${\tilde c}= c \ln 2$. For the pure, i.e., with no
disorder, antiferromagnetic spin-1/2 Heisenberg chain, which
presents conformal invariance with a central charge $c=1$, any
amount of disorder drives the ground state of this system to the
so-called Random Singlet Phase (RSP). This is a  gapless phase
with broken conformal symmetry  characterized by a collection of
singlet pairs of spins randomly distributed throughout the chain.
A schematic view of the RSP is displayed in Fig.~\ref{f1}. The RSP
appears not only for spin-1/2 chains, but also for higher spin
disordered chains. In the case of integer spin chains, the RSP
will usually arise whenever disorder is strong enough to close the
Haldane gap. From the point of view of quantum information
applications, the RSP is potentially useful for implementing
quantum channels with nearly perfect communication fidelity, since
it exhibits pairs of distant spins in highly entangled
states~\cite{Hoyos:06}.
One important open question for understanding the behavior of
entanglement in random quantum chains concerns how general the
logarithmic scaling of the von Neumann entropy is in these systems.
In this context, the RSP constitutes a rather interesting and
universal disordered ground state, appearing in a number of
half-integer and integer spin chains. In this letter, we
investigate the scaling of entanglement in the RSP of arbitrary
spin-$S$ chains. We find the logarithmic scaling holds in general
and furthermore it is associated with an {\it effective central
charge} $c_R = \ln(2S+1)$ which depends only on the value of the
spins in the chain.  These conclusions apply even if the pure
chain, i.e., with no disorder is described by a non-conformal
theory. Renormalization group calculations presented here fully
support our results.

\section{II - Renormalization group approach for random quantum spin
chains}

Renormalization group techniques provide a convenient
framework for the conceptualization of the quantum phases of
random spin chains. As concerns the RSP, it was first obtained
using a perturbative real-space renormalization group method
introduced by Ma, Dasgupta and Hu (MDH)~\cite{MDH1,MDH2}  to treat
the spin-1/2 REHAC. This approach was proven to be asymptotically
exact, which allowed for a fully characterization of the
properties of the RSP~\cite{Fisher:94}. Consider a chain of spins
described by the Heisenberg Hamiltonian
\begin{equation}
H = \sum_{i} J_i \overrightarrow{S}_i \cdot \overrightarrow{S}_{i+1},
\label{HH}
\end{equation}
where $\{\overrightarrow{S}_i\}$ is a set of spin-1/2 operators
and $\{J_i\}$ is a positive random variable obeying some
probability distribution $P(J)$. The original MDH method consists
in finding the strongest interaction $\Omega$ between a pair of
spins ($S_2$ and $S_3$ in Fig. 2a) and treating the couplings of
this pair with its neighbors ($J_1$ and $J_2$ in Fig. 2a) as a
perturbation. The  singlet formed by the spins coupled by the
strongest bond $\Omega$ is decimated away and an effective
interaction $J^\prime$ is perturbatively evaluated. By iteratively
applying this procedure, the low-energy behavior of the ground
state is obtained as a collection of singlet pairs formed over
arbitrary distances. This RSP is pictorially displayed in
Fig.~\ref{f1}.
\begin{figure}[th]
\centering {\includegraphics[angle=0,scale=0.4]{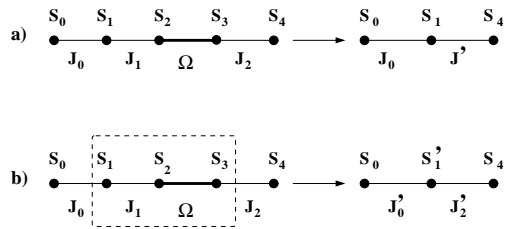}}
\caption{Modified MDH renormalization procedure for spin-$S$ chains.}
\label{f2}
\end{figure}

Unfortunately, when generalized to higher spins, this method, at least in its simplest version, revealed to
be ineffective. The reason is that, after the elimination procedure of the strongest bond $\Omega$, the effective
interaction $J^{\prime}$ may be greater than $\Omega$. Then, the problem becomes essentially non-perturbative
for arbitrary distributions of exchange interactions. For instance, considering
an arbitrary spin-$S$ REHAC, the renormalized coupling is given by the recursive relation~\cite{Boechat:96}
\begin{equation}
J^{\prime}=\frac{2}{3} S(S+1) \frac{J_1 J_2}{\Omega}.
\label{JPrimeH}
\end{equation}
Notice that, for $S \ge 1$, the renormalization factor is
$(2/3)S(S+1)>1$, resulting in a possible breakdown of perturbation
theory depending on the values of $J_1$ and $J_2$. In order to
solve this problem, a generalization of the MDH method was
proposed in Refs.~\cite{Saguia:02,Saguia:03}. This modified MDH
method consists in either of the following procedures shown in
Fig.~\ref{f2}. Taking the case of the Heisenberg chain as an
example, if the largest neighboring interaction to $\Omega$, say
$J_1$, is $J_1 < 3/(2S(S+1)) \Omega$, then we eliminate the
strongest coupled pair obtaining an effective interaction between
the neighbors to this pair which is given by Eq.~(\ref{JPrimeH})
(see Fig.~2a). This new effective interaction is always smaller
than those eliminated. Now suppose $J_1 > J_2$ and $J_1 >
3/(2S(S+1)) \Omega$. In this case, using Eq.~(\ref{JPrimeH}) would
give rise to an interaction larger then those eliminated. In order
to avoid that, we consider the {\em trio} of spins-$S$ coupled by
the two strongest interactions of the trio, $J_1$ and $\Omega$ and
solve it exactly (see Fig.~2b). The trio of spins is then
substituted by one effective spin-S interacting with its neighbors
through new renormalized interactions obtained by degenerate
perturbation theory for the ground state of the trio. This method
has been successfully  applied to  investigate the quantum phase
diagram of the spin-1~\cite{Saguia:02} and
spin-3/2~\cite{Saguia:03} REHACs. As we show below, it turns out
to be also essential for the computation of block entanglement in
high spin chains.

\section{III - Entanglement entropy in the RSP}

Let us consider the general structure of the RSP shown schematically in Fig.~\ref{f1}.
The ground state of the system consists of pairs of spins coupled
into singlets over arbitrary distances. These can be represented
at zeroth order by the tensor product $|\psi^{(0)}\rangle =
|\psi^{-}_{i_1 i_2}\rangle \otimes |\psi^{-}_{i_3 i_4}\rangle
\dots \otimes |\psi^{-}_{i_{n} i_{n+1}}\rangle$, where
$|\psi^{-}_{i_k i_{k+1}}\rangle$ denotes the singlet state
$(1/\sqrt{2})(|\uparrow \downarrow \rangle - |\downarrow \uparrow
\rangle )$ for spins at arbitrary distant sites labelled by
integer numbers $i_k$ and $i_{k+1}$. In order to evaluate
entanglement, we then observe that von Neumann entropy of a tensor
product $\rho \otimes \sigma$ for density matrices $\rho$ and
$\sigma$ is simply ${\cal S}(\rho \otimes \sigma) = {\cal S}(\rho)
+ {\cal S}(\sigma)$~\cite{Nielsen:book}. Therefore, the
entanglement entropy of a block of length $L$ can be obtained by
counting singlet pairs which cross the boundary of the block. This
yields
\begin{equation}
{\cal S}(L) = {\cal S}_p \, \langle N_S (L) \rangle,
\label{EntangAverage}
\end{equation}
where ${\cal S}_p$ is the entanglement entropy of a pair of spins
and $\langle N_S (L) \rangle$ is the configurational average of
the number of singlets connecting the two blocks. For a pair of
spin-$S$ particles in a singlet state, von Neumann entropy of the
pair is maximal and given by ${\cal S}_p = \log_2 \left( 2S+1
\right)$. Concerning $\langle N_S (L) \rangle$, it has been
evaluated analytically for the spin-1/2 random Heisenberg
antiferromagnetic chain in Ref.~\cite{Refael:04} via counting of
singlets from the explicit solution of the renormalization group
flow equation for the distribution of couplings. The value
obtained for $\langle N_S (L) \rangle$ reads
\begin{equation}
\langle N_S (L) \rangle = \frac{\ln 2}{3} \log_2 L + k,
\label{N}
\end{equation}
with $k$ a non-universal constant. Although Eq.~(\ref{N}) was derived for the specific
case of spin-1/2,
it should actually hold for any
spin-$S$ chain in the RSP. The reason is that the average number of singlet pairs should be the
same for any ground state which is a completely random collection of singlets along the chain.
Indeed, $\langle N_S (L) \rangle$ is a number which should be a general property of the structure
of the RSP, being independent of the magnitude $S$ of the spins composing the chain.
Therefore, by inserting Eq.~(\ref{N}) into Eq.~(\ref{EntangAverage}), we obtain for any
random quantum spin-$S$ chain in the RSP the following relationship:
\begin{equation}
{\cal S}(L) = \frac{c_{R}}{3}\, \log_2 L + k \, ,
\label{ScalingRandom}
\end{equation}
where
\begin{equation}
c_{R}= \ln 2 \,  \log_2 \left( 2S+1 \right)  = \ln \left( 2S+1 \right).
\label{cR}
\end{equation}
Notice that $c_R$ depends only on the value of $S$, which is therefore the
key object to set the effective central charge of random spin chains in the RSP.
Notice that Eq.~(\ref{cR}) does not involve any central charge of pure (non-disordered)
models, since we do not have a central charge associated with integer spin chains
exhibiting a Haldane gap.
Let us turn now to the explicit evaluation of entanglement entropy for the spin-$1$
and spin-$3/2$ REHACs via numerical implementation of the renormalization
group equations. As it will be seen in the next sections, this will provide a full
support to our scaling law for ${\cal S}(L)$.

\subsection{III.1 - Spin-1 and spin-3/2 REHACs}

\begin{figure}[th]
\centering {\includegraphics[angle=0,scale=0.35]{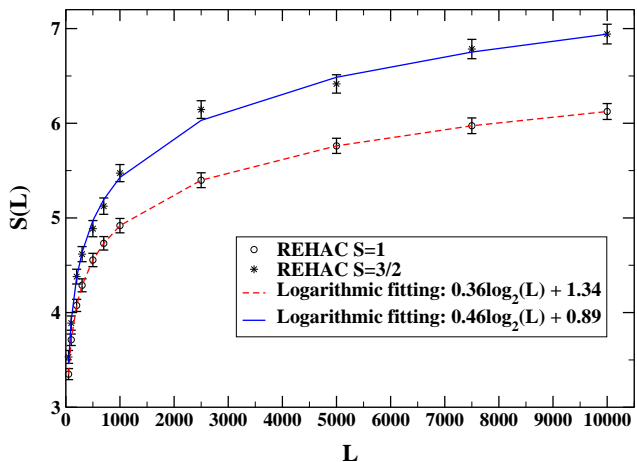}}
\caption{Logarithmic scaling of entanglement entropy for the spin-1 and spin-3/2 REHACs in the RSP.
Blocks of spins up to 10,000 sites are considered in a chain initially with 200,000 sites. The logarithmic
fittings shown is the best curve fitting numerically obtained.}
\label{f3}
\end{figure}

We apply now the generalized MDH method to discuss entanglement in the spin-1 and
spin-3/2 Heisenberg chains. The Hamiltonian for these chains will
be given by Eq.~(\ref{HH}), but with the set
$\{\overrightarrow{S}_i\}$ denoting now spin-1 and spin-3/2
operators, respectively. Decimation of the chain follows the
procedure previously described, with the renormalized couplings
given by Eq.~(\ref{JPrimeH}). Then, block entanglement can be
numerically computed in the following way. If a singlet is
decimated and the spins composing the singlet are in different
blocks, this singlet adds $\log_2 (2S+1)$ to the von Neumann
entropy. On the other hand, in the case of a {\em trio}
elimination, we have seen that one effective spin is returned to
the chain. Then, if the spins composing the {\em trio} are in
different blocks, the effective spin is introduced, by a majority
vote strategy, in the block contributing with more spins to the
{\em trio},  such that, each block contributes with one spin for
the elimination process. At the time the effective spin forms a
singlet with another spin (either effective or not), entanglement
will then be  counted as in the singlet case previously described.
Hence, notice that von Neumann entropy is counted only when a
singlet decimation occurs. This is somewhat similar to the
procedure adopted in Ref.~\cite{Refael:04} for the transverse
field Ising model, where entanglement was counted only for
effective spins decimated by the transverse field. In our
numerical procedure, we averaged entanglement over 1000 random
coupling configurations following a power-law probability
distribution of couplings $J_i$ given by $P(J) \sim J^{-0.8}$, for
which, {\em trio} renormalizations are in a negligible number in
the RSP 
\footnote{Power-law probability distributions are numerically convenient (but not fundamental) to 
make evident the scaling of entropy in the RSP, since {\em trio} renormalizations get rather negligible 
in this scenario. Indeed, our results were checked for a number of negative exponents, 
even for very small ones and also zero (square distribution case).}. 
Numerical analysis of entanglement as a function of the
length of the block is plotted in Fig.~\ref{f3}, where we
considered blocks of spins up to 10000 sites in a chain initially
with 200000 sites. Notice that: (i) block entanglement scales
logarithmically with the length $L$ of the block; (ii) the
logarithmic scaling follows our Eq.~(\ref{ScalingRandom}), with
$c_R= \ln 3$ for spin-1 and $c_R= \ln 4$ for spin-3/2. The case of
spin-1 is particularly remarkable, since the scaling is governed
by an effective central charge although the model does not exhibit
conformal invariance even in the limit of vanishing disorder.


\subsection{III.2 - Random biquadratic spin-1 chain}

\begin{figure}[th]
\centering {\includegraphics[angle=0,scale=0.35]{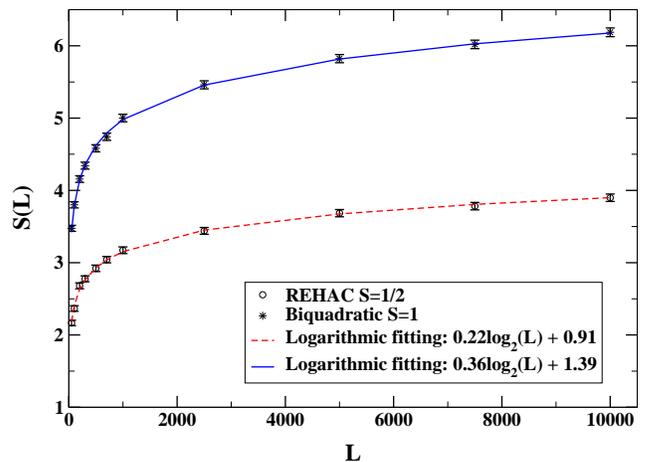}}
\caption{Logarithmic scaling of entanglement entropy for the random biquadratic spin-1 chain and 
the spin-1/2 REHAC in the RSP.
Blocks of spins up to 10,000 sites are considered in a chain initially with 200,000 sites.
The logarithmic fittings shown is the best curve fitting numerically obtained.}
\label{f4}
\end{figure}

As shown by Eq.~(\ref{cR}), scaling in the RSP depends only on the magnitude
of the spins composing the chain, being independent of the
particular properties of the model. Indeed, this can be
illustrated by considering the spin-1 biquadratic chain, whose
Hamiltonian is given by
\begin{equation}
H = \sum_{i} \Delta_i \left( \overrightarrow{S}_i \cdot \overrightarrow{S}_{i+1} \right) ^2,
\label{HB}
\end{equation}
where $\{\overrightarrow{S}_i\}$ is a set of spin-1 operators and $\{\Delta_i\}$ is a positive random
variable obeying some probability distribution $P(J)$. Application of the generalized MDH procedure
here results only in the formation of singlets, with the renormalized exchange coupling reading~\cite{Boechat:96}
\begin{equation}
\Delta^{\prime}=\frac{2}{9} \frac{\Delta_1 \Delta_2}{\Omega},
\label{JPrimeB}
\end{equation}
where $\Delta_1$ and $\Delta_2$ are the neighbors to the strongest
bond. Notice that {\em trio} renormalizations are completely
absent here, since the renormalized couplings generated are always
smaller than the energy scale $\Omega$. Similar behavior occurs
for the spin-1/2 Heisenberg chain, with $J^\prime = (1/2) J_1 J_2
/ \Omega$ and, therefore, the original MDH method can be
straightforwardly applied. In this simpler situation, entanglement
between a block of length $L$ and the rest of the chain can be
computed by exclusively singlet renormalizations (there is no
possibility of {\em trio} renormalizations). Block entanglements
for both the biquadratic spin-1 chain and the spin-1/2 Heisenberg
chain are shown in Fig.~\ref{f4}. As before, we considered blocks
of spins up to 10000 sites in a chain initially with 200000 sites,
with entanglement averaged over 1000 random coupling
configurations following a power-law probability distribution of
couplings $J_i$ given by $P(J) \sim J^{-0.8}$. For the spin-1/2
Heisenberg chain, the result for the scaling is in agreement with
Ref.~\cite{Refael:04} and reproduces Eq.~(\ref{EntangAverage}).
Moreover, observe that both the spin-1 REHAC in Fig.~\ref{f3} and
the random biquadratic chain exhibit the same {\it effective
central charge} $c_R$, in complete agreement with Eq.~(\ref{cR}).

\section{IV - conclusion}
In summary, we have characterized the logarithmic scaling of
entanglement entropy in the RSP of random quantum chains of
general spin-$S$. In particular, we have shown that an {\it
effective central charge} governing the scaling can be defined,
which depends exclusively on the magnitude $S$ of the spin, as
given by Eq.~(\ref{cR}). In order to give support to these
conclusions, we evaluated entanglement for different spin-1 and
spin-3/2 chains using a real space renormalization group approach.
These results are encouraging for the pursuit of a complete
characterization of entanglement entropy in general random
critical spin-$S$ chains. In particular, characterization of the
complete phase diagram, i.e., for different degrees of disorder of
spin-1 and spin-3/2 REHACS, is challenging. This topic is left for
a future investigation.

\section{Acknowledgments}
This work was supported by the Brazilian agencies CAPES (A.S.), CNPq (M.S.S., B.B., and M.A.C.),  and
FAPERJ (M.S.S. and M.A.C.).

\end{document}